\documentclass[reprint,
onecolumn, 
superscriptaddress,
showkeys,
 amsmath,amssymb,
 aps,
 prf,
]{revtex4-2}
\usepackage{graphicx}
\usepackage{amsmath}
\usepackage{dcolumn}
\usepackage{bm}
\usepackage[dvipsnames]{xcolor}
\usepackage{caption}
\usepackage{soul}
\usepackage{eucal}

\usepackage{subcaption}
\begin{document}

\title{Taylor dispersion in a soft tube}
\author{Aditya Jha}
	\email{aj756@cam.ac.uk}
 \affiliation{TCM Group, Cavendish Laboratory, University of Cambridge, Cambridge CB3 0US, United Kingdom.}
  \affiliation{Univ. Bordeaux, CNRS, LOMA, UMR 5798, F-33405 Talence, France.}
 \author{Masoodah Gunny}
 \affiliation{Gulliver UMR 7083 CNRS, ESPCI–PSL, 10 rue Vauquelin, 75005 Paris, France.}
  \affiliation{IPGG, 6 rue Jean-Calvin, 75005 Paris, France.}
 \author{Joshua D. McGraw}
 \affiliation{Gulliver UMR 7083 CNRS, ESPCI–PSL, 10 rue Vauquelin, 75005 Paris, France.}
 \affiliation{IPGG, 6 rue Jean-Calvin, 75005 Paris, France.}
\author{Yacine Amarouchene}
 \affiliation{Univ. Bordeaux, CNRS, LOMA, UMR 5798, F-33405 Talence, France.}
\author{Thomas Salez}
	\email{thomas.salez@cnrs.fr}
 \affiliation{Univ. Bordeaux, CNRS, LOMA, UMR 5798, F-33405 Talence, France.} 
\date{\today}

\begin{abstract}
Diffusion of a solute along a tube is enhanced by hydrodynamic flow, a phenomenon known as Taylor dispersion. In microfluidic applications, the compliance of the tube boundaries modifies the hydrodynamic flow and thus solutal transport. Here, we develop the theory of solutal dispersion in a soft, axisymmetric tube where the tube walls respond to the hydrodynamic pressure through a Winkler response. By deriving the modified macro-transport equation for the solutal concentration dynamics based on multiple-time-scale analysis, we explore the influence of softness on solutal transport for steady and pulsatile configurations. Our main finding is that softness enhances the effective advection velocity and dispersion coefficient, which might have practical implication in biology and microfluidic technology.
\end{abstract}

\maketitle

\section{\label{sec:intro}Introduction}
Transport in microtubes plays a fundamental role in living systems. Our understanding of these small-scale flows in the past few decades has been revolutionised with the advent of improved microfabrication technology aiding microfluidic designs~\cite{whitesides2006origins,stone2001microfluidics}. Specifically, a myriad of important biophysical processes including cellular signalling~\cite{alim2017mechanism} and chemical reactions~\cite{shapiro1986taylor} can be understood by carefully looking at the dynamics of particles moving inside microtubes. With the emergence and the increased use of soft materials including polymeric gels and elastomers, the tube elasticity and its coupling with the fluid flow has become increasingly important. Flows in deformable tubes have been intensively studied over the past decade both experimentally~\cite{guyard2022elastohydrodynamic,boyko2020interfacial,boyko2020nonuniform} and theoretically~\cite{christov2021soft,christov2018flow,huang2025oscillatory,martinez2020start,anand2020transient,wang2019theory,rade2025theory} with the aim of characterising the flow-deformation coupling. These studies have characterising the changes to the volumetric flow rate and/or pressure gradients in the tube owing to the softness of the walls. Recent studies have highlighted how this coupling can lead to rectified flow controlled by the wall elasticity~\cite{zhang2024elasto,rade2025theory}. The effect of the non-Newtonian nature of the fluid in such tubes has also been studied~\cite{boyko2023non,chun2025experimental}, while pulsatile flows have been suggested as a mechanism for characterising the elasticity of the surrounding walls~\cite{pande2023oscillatory,rade2025theory}. 

The transport of solutes in soft microfluidic tubes, where the flow is modified by its coupling with the elasticity of the wall, is expected to be fundamentally different from that in rigid tubes. In rigid tubes, the diffusive dynamics of the particles is coupled with the advection by the flow, a classical phenomenon called Taylor dispersion or Taylor-Aris dispersion~\cite{taylor1953dispersion,aris1956dispersion,chatwin1970approach,chatwin1977initial,frankel1989foundations,vedel2012transient,vedel2014time,guyard2021near}. Many studies have extended the scope to other scenarios including oscillating shear flows~\cite{schmidt2005effect}, the effects of chemical reactions and/or absorption at the wall~\cite{sankarasubramanian1973unsteady,smith1983effect,barton1984asymptotic,shapiro1986taylor,datta2008dispersion,levesque2012taylor,Vilquin2021,vilquin2023nanoparticle}, boundary thermal fluctuations~\cite{marbach2018transport,sarfati2021enhanced}, non-Newtonian fluid flows~\cite{sharp1993shear,rana2016solute,rana2016unsteady}, and complex geometries~\cite{dorfman2002generalized,dutta2006effect,dutta2001dispersion,aminian2016boundaries}. Recent studies have highlighted the influence of a slowly-varying profile along the tube~\cite{chang2023taylor}, and of active pumping of the tube walls~\cite{marbach2019active}, that lead to a non-trivial dynamics and are useful in controlling the dispersion of solutes in microtubes. Other studies have further addressed active particles and their dispersion in such flows, both theoretically~\cite{ezhilan2015transport,jiang2019dispersion,peng2020upstream,jiang2021transient}, and experimentally~\cite{lagoin2025enhanced}, in an attempt to understand living systems in complex environnements in more detail. Even though such a diverse set of theory and modelling has been done, the dynamics of solutes and the modified dispersion when the flow couples to the elasticity of the wall remains an open question. Along the same line, the general coupling between elastohydrodynamics and Brownian motion is a recent research topic~\cite{bickel2006brownian,daddi2016particle,Fares2024}. 

In this study, we explore the modification induced to Taylor-Aris dispersion when the flow couples to the elasticity of the boundaries for constant and pulsatile pressure at the inlet of an axisymmetric tube. We use the theory of flows in elastic microtubes, by incorporating the elasticity through a Winkler model. The solute transport is described by multiple-time-scale analysis~\cite{ng2006dispersion,chu2019dispersion} that allows us to describe the long-time dynamics of the cross-sectionally-averaged concentration. We then characterise the changes induced to the advection velocity and dispersion coefficient due to tube softness. In order to illustrate the main softness-induced effects for practical purposes, we end our discussion by computing the spatiotemporal evolution of a concentration peak of solute under steady flow~\cite{danckwerts1995continuous,trachsel2005measurement,rodrigues2021residence}. 


\section{Theory}
\begin{figure}[h]
\includegraphics[width=8cm]{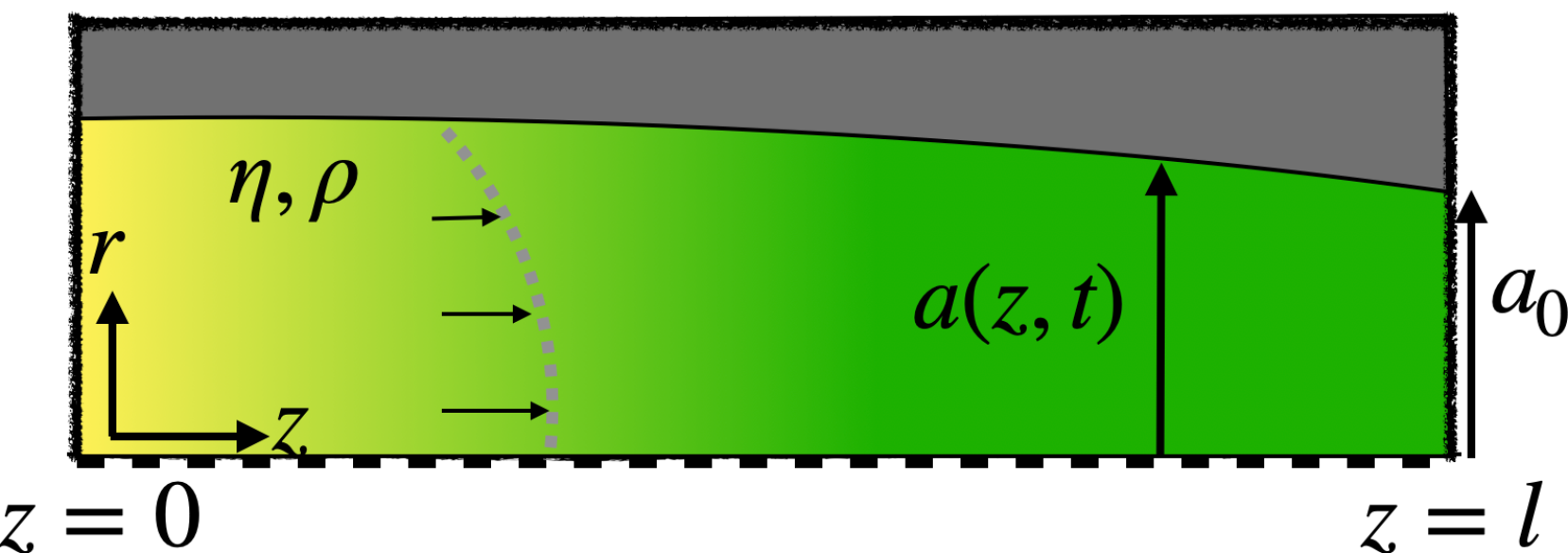}
\caption{Schematic of the system under study. We consider a flow of a viscous fluid, with viscosity $\eta$ and density $\rho$, in an elastic, axisymmetric tube. We denote the radial coordinate by $r$, $r=0$ corresponding to the revolution axis, and the axial coordinate by $z$. The length of the tube is $l$ and the radius at the exit of the tube is $a_0$. The local radius of the tube is denoted by $a(z,t)$. The solute concentration is highlighted by the color in the channel transitioning from high concentration (yellow) to low concentration (green).}
\label{fig_1}
\end{figure}

We consider an axisymmetric tube with a wall made of an elastic material, whose elasticity can be modelled as a bed of independent springs, leading to a Winkler law for deformation under applied pressure. A schematic of the system is presented in Fig.~\ref{fig_1}, where $a(z,t)$ denotes the local radius of the tube that varies along the longitudinal direction $z$ and time $t$. Here, $a_0$ denotes the radius at the exit and $l$ denotes the length of the tube. The fluid is assumed to be Newtonian with viscosity $\eta$ and density $\rho$. We consider the hydrodynamic pressure (in excess to the atmospheric one) at the inlet, of amplitude $p_0$, to be either steady, or oscillatory with angular frequency $\omega$, and the outlet to be open to the atmosphere implying $p(z = l) = 0$. The fluid contains a solute, with a diffusion constant $D$ and a typical concentration $c_0$.

\subsection{Governing equations}
The flow in the tube is governed by the incompressible Navier-Stokes equations with the solute transport described by an advection-diffusion equation. The problem is assumed to be axisymmetric. Hence, the conservation equations for fluid volume, momentum and solute matter are: 
\begin{align}
    \frac{\partial u}{\partial z}+\frac{1}{r}\frac{\partial}{\partial r}\left(rv\right) &= 0~,\label{eq:incompressibility_dim}\\
    \rho\frac{\partial v}{\partial t}+\rho\left(u\frac{\partial v}{\partial z}+v\frac{\partial v}{\partial r}\right)&=-\frac{\partial p}{\partial r}+ \eta\left[\frac{\partial^2 v}{\partial z^2}+\frac{1}{r}\frac{\partial}{\partial r}\left(r\frac{\partial v}{\partial r}\right)-\frac{v}{r^2}\right]~,\\
    \rho\frac{\partial u}{\partial t}+\rho\left(u\frac{\partial u}{\partial z}+v\frac{\partial u}{\partial r}\right)&=-\frac{\partial p}{\partial z} + \eta\left[\frac{\partial^2 u}{\partial z^2}+\frac{1}{r}\frac{\partial}{\partial r}\left(r\frac{\partial u}{\partial r}\right)\right]~,\\
    \frac{\partial c}{\partial t}+u\frac{\partial c}{\partial z}+v\frac{\partial c}{\partial r} & = D\left[\frac{\partial^2 c}{\partial z^2}+\frac{1}{r}\frac{\partial}{\partial r}\left(r\frac{\partial c}{\partial r}\right)\right]\ ,
\end{align}
where $v(r,z,t)$, $u(r,z,t)$, $p(r,z,t)$, and $c(r,z,t)$ are the radial velocity, axial velocity, hydrodynamic pressure, and concentration fields, respectively. We further assume a no-slip boundary condition at the wall, \textit{i.e.} $u(r=a,z,t)=0$. Fluid volume conservation implies:
\begin{align}
    \frac{\partial (\pi a^2)}{\partial t}+\frac{\partial q}{\partial z} = 0~, 
\end{align}
with the cross-sectional fluid flux defined by:
\begin{align}
    q = 2\pi \int_0^a\textrm{d}r\, r\, u~.
\end{align}

For closure, the constitutive elastic relation between pressure and wall deformation needs to be specified. To this purpose, we assume a Winkler response, \textit{i.e.} a linear and local elastic response where the deformation is proportional to the  pressure~\cite{gervais2006flow,dendukuri2007stop,alim2017mechanism,wang2019theory} (evaluated at $r = a$), as:
\begin{align}
    a =  a_0+ kp~, 
    \label{wink}
\end{align}
where $k$ is a coefficient (with dimension of a volume per unit force) characterizing the compliance of the wall. 

We look for solutions to the equations above under the lubrication approximation, in particular, where $\epsilon = a_0/l\ll 1$. At this point, it becomes relevant to look at the various time scales involved in the problem. First, the diffusion time across the tube radius is given by the Taylor time~\cite{fife1975dispersion,ng2006dispersion,chu2019dispersion}:
\begin{align}
    \tau_0 = \frac{a_0^2}{D} ~.
\end{align}
Second, the advection time across the length of the tube is given by:  
\begin{align}
     \tau_1 = \frac{l}{U_0}= \frac{ \tau_0}{\text{Pe}\,\epsilon}~,
\end{align}
where $U_0=p_0a_0^2/(l\eta)$ denotes the characteristic fluid velocity scale along the axial direction, and where we introduced the P\'eclet number $\text{Pe} = U_0a_0/D$, assumed to be $O(1)$. In the specific case of oscillatory flows, we also assume $ \tau_1$ to be comparable to the flow oscillation period $2\pi/\omega$. Third, the diffusion time scale across the length of the tube is given by: 
\begin{align}
     \tau_2 = \frac{l^2}{D}= \frac{ \tau_0}{\epsilon^2}~.
\end{align}
One thus has the time-scale hierarchy $ \tau_0\ll \tau_1\ll  \tau_2$. In the following, we will ignore the fast dynamics happening at time scales comparable to $ \tau_0$, since we aim at understanding the concentration evolution on much longer times~\cite{ng2006dispersion} and these dynamics govern the transient evolution of the solute profile before the long-time behaviour sets in. Note that we also neglected the cross-channel viscous time scale $a_0^2\rho/\eta$, and any viscoelastic or acoustic time scale for the response of the elastic material, as we assumed these to be even smaller than $ \tau_0$.

Let us now non-dimensionalize the problem through:
\begin{align}
z& = lZ~,   &r& = a_0R~,  & u & =U_0U(R,Z,T)~,   &v & =\epsilon U_0V(R,Z,T)~,\nonumber\\  a(z,t)& = a_0A(Z,T)~,& p(r,z,t)&=p_0P(R,Z,T)~, & t& =\frac{2\pi}{\omega}T~,&c(r,z,t)&=c_0C(R,Z,T)~.
\end{align}
With these scalings, the dimensionless versions of the governing equations read: 
\begin{align}
    \frac{\partial U}{\partial Z}+\frac{1}{R}\frac{\partial }{\partial R}\left(RV\right) &= 0~,\\
    \frac{\epsilon^2\text{Wo}^2}{2\pi}\frac{\partial V}{\partial T}+\epsilon^3 \text{Re}\left(U\frac{\partial V}{\partial Z}+V\frac{\partial V}{\partial R}\right) &=-\frac{\partial P}{\partial R}+\epsilon^4\frac{\partial^2 V}{\partial Z^2}+\epsilon^2\frac{1}{R}\frac{\partial}{\partial R}\left(R\frac{\partial V}{\partial R}\right)-\epsilon^2\frac{V}{R^2}~,\\
    \frac{\text{Wo}^2}{2\pi}\frac{\partial U}{\partial T}+\epsilon \text{Re}\left(U\frac{\partial U}{\partial Z}+V\frac{\partial U}{\partial R}\right) &=-\frac{\partial P}{\partial Z}+\epsilon^2\frac{\partial^2 U}{\partial Z^2}+\frac{1}{R}\frac{\partial}{\partial R}\left(R\frac{\partial U}{\partial R}\right)~,\\
    \phi\epsilon\frac{\partial C}{\partial T}+\epsilon \text{Pe}\left(U\frac{\partial C}{\partial Z}+V\frac{\partial C}{\partial R}\right) &= \frac{1}{R}\frac{\partial }{\partial R}\left(R\frac{\partial C}{\partial R}\right)+\epsilon^2\frac{\partial^2 C}{\partial Z^2}~,\label{eq:conc-eqn-ND}\\
     \text{St}\,A\,\frac{\partial A}{\partial T}+\frac{\partial }{\partial Z}\int_0^A\text{d}R\,R\,U&=0~,   \\
A &=  1+{\kappa P}~, \label{eq:ND_height_pressure}
\end{align}
where we introduced the dimensionless parameter $\phi= \omega l a_0/(2\pi D)$, and the dimensionless compliance $\kappa=kp_0/a_0$, and where $\text{Re} = \epsilon \rho a_0^2p_0/\eta^2$, $\text{Wo} = a_0\sqrt{\rho \omega/\eta}$, and $\text{St} = \omega \eta/(2\pi p_0\epsilon^2)$, denote the Reynolds, Womersley and Strouhal numbers, respectively. Note that the latter corresponds to the dimensionless expression of $\tau_1$. These dimensionless numbers characterise the relative strengths of either the convective inertia ($\textrm{Re}$) or the unsteady effects ($\textrm{Wo \& St}$) with respect to the viscous forces. They are related through $\textrm{Wo}^2 = 2\pi \textrm{St}(\epsilon \textrm{Re})$. The notation for all the dimensionless numbers are summarised in Tab.~\ref{tab:parameter-table} including their typical values.

In the following, we focus on the solutions of the governing equations in the regime where $\epsilon \ll1$, $\epsilon \text{Re}\ll1$, $\epsilon^2\text{Wo}^2/(2\pi)\ll1$, and $\text{Pe}\sim O(1)$. In the specific case of oscillatory flows, we further assume $\phi\sim O(1)$ and $\text{St}\sim O(1)$. Since the flow is independent of the solute concentration dynamics, we address these separately, starting with the former. 

\begin{table}
  \begin{center}
\def~{\hphantom{0}}
  \begin{tabular}{lcc}
      $\textrm{Quantity} $  & $\textrm{Notation}$ & $\textrm{Typical Value}$  \\ \hline\hline 
        $\textrm{Aspect ratio of the tube} $ & $\epsilon = a_0/l $ & $0.001 - 0.01$\\
        $\textrm{Reynolds number}$ & \text{Re} = $\epsilon \rho a_0^2p_0/\eta^2$ & $0.1 - 100$ \\
       $\textrm{Wommersley number}$  & $\text{Wo} = a_0\sqrt{\rho \omega/\eta}$ & $0.1 - 10$ \\
        $\textrm{Strouhal number}$ & $\text{St} = \omega \eta/(2\pi p_0\epsilon^2)$ & $10^{-9} - 10^{-6}$ \\
       $\textrm{Compliance}$ & $\kappa=kp_0/a_0 $ & $0.001 - 0.01$ \\
  \end{tabular}
  \caption{Key dimensionless parameters and associated ranges.}
    \label{tab:parameter-table}
  \end{center}
\end{table}

\subsection{Flow profile}
\label{floprof}
The leading-order governing equation for the fluid velocity field reads
\begin{align}
    \frac{\text{Wo}^2}{2\pi}\frac{\partial U}{\partial T}&=-\frac{\partial P}{\partial Z}+\frac{1}{R}\frac{\partial}{\partial R}\left(R\frac{\partial U}{\partial R}\right)~,
\end{align}
together with $\partial P/\partial R=0$, and the no-slip boundary condition at the wall $U(R=A,Z,T)=0$.

The steady case (labeled with the ``s" subscript) can be described for the flow part by setting $\text{Wo}=0$ and $\textrm{St}=0$, as well as $T=0$ (or equivalently dropping the $T$ variable). Hence, the steady fluid velocity field in the axial direction reads:
\begin{align}
    U_{\text{s}}(R,Z) = \left(\frac{R^2-A^2}{4}\right)\frac{\text{d}P_{\text{s}}}{\text{d}Z}~,\label{eq:steady-axial-velocity-general}
\end{align}
with $A=1+\kappa P_{\text{s}}$.
The mean axial velocity across the cross section is thus given by: 
\begin{align}
    \langle U_{\text{s}}\rangle(Z) = \frac{2}{A^2}\int_0^A\textrm{d}R\,R\,U_{\text{s}} = -\frac{A^2}{8}\frac{\text{d}P_{\text{s}}}{\text{d}Z}~,
    \label{mina}
\end{align}
where $\langle \cdot \rangle$ denotes cross-sectional averaging. Furthermore, the steady cross-sectional fluid flux is invariant along the tube axis, leading to: 
\begin{align}
 \frac{\text{d}}{\text{d}Z}\left[(1+\kappa P_{\text{s}})^4\frac{\text{d}P_{\text{s}}}{\text{d}Z}\right]=0\ .
    \label{eq:ND_flux_conservation_steady}
\end{align}
The latter equation, along with the boundary conditions $P_{\text{s}}(Z = 0) = 1$ and $P_{\text{s}}(Z = 1) = 0$, leads to the hydrodynamic pressure field~\cite{guyard2022elastohydrodynamic,pande2023oscillatory}: 
\begin{align}
      P_{\text{s}}(Z) = \frac{1}{\kappa}\left(\left\{(1-Z)[(1+\kappa)^5-1]+1\right\}^{1/5}-1\right)~.
      \label{hydrop}
\end{align}

For the oscillatory case (labeled with the ``p" subscript), we assume the axial pressure gradient to be in the form $\partial P_{\text{p}}/\partial Z = -\mathcal{R} \{G(Z)\textrm{e}^{i2\pi T}\}$, where $\mathcal{R}(\cdot)$ denotes the real part and $G(Z)$ is a complex amplitude to be determined. Correspondingly, the axial velocity field is written as $U_{\text{p}} = \mathcal{R}\{\tilde{U}_{\text{p}}\}$, with $\tilde{U}_{\text{p}}=H(R,Z)\textrm{e}^{i2\pi T}$, where $H(R,Z)$ is a complex amplitude to be determined. Doing so, one gets~\cite{womersley1957oscillatory,pande2023oscillatory}: 
\begin{align}
    H(R,Z) = \frac{G(Z)}{i\text{Wo}^2}\left[1-\frac{J_0(i^{3/2}R\text{Wo})}{J_0(i^{3/2}A\text{Wo})}\right]~,
\end{align}
which thus leads to:
\begin{align}
    U_{\text{p}}(R,Z,T) = \mathcal{R}\left\{\frac{i}{\text{Wo}^2}\left[1-\frac{J_0(i^{3/2}R\text{Wo})}{J_0(i^{3/2}A\text{Wo})}\right]\frac{\partial \tilde{P}_{\text{p}}}{\partial Z}\right\}~,\label{eq:periodic-axial-velocity-general}
\end{align}
where we introduced the complex pressure field $\tilde{P}_{\text{p}}$, related to the real one through $P_{\text{p}} = \mathcal{R}\{\tilde{P}_{\text{p}}\}$, and the Bessel function of the first kind of order zero $J_0$. Then, fluid volume conservation implies:
\begin{align}
    \text{St}\,A\, \frac{\partial A}{\partial T} =\mathcal{R}\left\{\frac{\partial }{\partial Z}\left[\frac{iA^2}{2\text{Wo}^2}\frac{\partial \tilde{P}_{\text{p}}}{\partial Z}\frac{J_2(i^{3/2}A\text{Wo})}{J_0(i^{3/2}A\text{Wo})}\right]\right\}~,
    \label{eq:ND_flux_conservation_oscillatory}
\end{align}
where $J_2$ is the Bessel function of the first kind of order two. We stress that this methodology of capturing the flow profile in a deforming channel under pulsatile flow is based on the von K\'arm\'an-Polhausen approximation~\cite{pande2023oscillatory,stewart2009local}. Equation~(\ref{eq:ND_flux_conservation_oscillatory}) together with Eq.~(\ref{eq:ND_height_pressure}), \textit{i.e.} $A=1+\kappa \mathcal{R}\{\tilde{P}_{\text{p}}\}$, and the boundary conditions: 
\begin{align}
    \tilde{P}_{\text{p}}(Z = 0,T) &= \textrm{e}^{2\pi iT}\ ,\\
    \tilde{P}_{\text{p}}(Z =1,T) &= 0\ , 
\end{align}
form a closed set of equations for the pressure that can be solved numerically. Interestingly, due to the elastic coupling in Eq.~(\ref{eq:ND_height_pressure}), Eq.~(\ref{eq:ND_flux_conservation_oscillatory}) is non-linear in the pressure field, and thus expected to generate mode-coupling effects. At small elastic compliance, we can approach the problem analytically with a perturbation expansion in $\kappa$~\cite{pande2023oscillatory}, as provided in the Appendix. 

\subsection{Solute transport analysis}
We employ multiple-time-scale analysis~\cite{fife1975dispersion,ng2006dispersion,chu2019dispersion} to arrive at the evolution of the cross-sectionally-averaged concentration. Although there are other methods that have been commonly used to study dispersion, including invariant-manifold analysis, method of moments, and others~\cite{aris1956dispersion,vedel2012transient,marbach2018transport,mercer1990centre,mercer1994complete}, we focus on the multiple-time-scale analysis as it leads to the governing equation for the cross-sectionally averaged concentration dynamics in a way that is convenient for both steady and oscillatory flows. The solute dynamics is given by Eq.~(\ref{eq:conc-eqn-ND}), together with the solutal no-flux boundary condition at the wall:
\begin{align}
  \frac{\partial C}{\partial R}\bigg|_{R=A} = \epsilon^2\frac{\partial A}{\partial Z}\frac{\partial C}{\partial Z}\bigg|_{R=A}~.   
  \label{cbc}
\end{align}
Following standard multiple-time-scale analysis~\cite{bender1999advanced,holmes2012introduction}, the time derivative is first expanded as: 
\begin{align}
    \frac{\partial}{\partial T} = \frac{\partial}{\partial T_1}+\epsilon\frac{\partial}{\partial T_2}~,
    \label{t1t2}
\end{align}
where $T_1$ is a ``fast" time variable describing dynamics at $T\sim\omega\tau_1/(2\pi)\sim O(1)$, and $T_2$ is a ``slow" time variable describing dynamics at $T\sim\omega\tau_2/(2\pi)\sim O(1/\epsilon)$. In addition, the concentration field is decomposed as~\cite{fife1975dispersion,ng2006dispersion,chu2019dispersion}: 
\begin{align}
    C(R,Z,T) = \sum_{n=0}^\infty\epsilon^nC^{(n)}(R,Z,T_1,T_2)~.
\end{align}
Substituting the above decompositions in Eqs.~(\ref{eq:conc-eqn-ND}) and~(\ref{cbc}), we arrive at the governing equations at different orders in the perturbation parameter $\epsilon$. Below, we address these successively up to second order. 

\subsubsection{Leading-order analysis}
At $O(1)$, the governing equation reads: 
\begin{align}
    \frac{1}{R}\frac{\partial}{\partial R}\left(R\frac{\partial C^{(0)}}{\partial R }\right) = 0~,
\end{align}
which must satisfy the boundary condition: 
\begin{align}
    \frac{\partial C^{(0)}}{\partial R}\bigg|_{R=A} = 0~.
\end{align}
The solution of these two equations indicates that $C^{(0)}$ is independent of $R$. Since the $O(1)$ solution corresponds in our case to dynamics at times $T$ much larger than the dimensionless Taylor time $\omega\tau_0/(2\pi)$, it is natural 
to identify $C^{(0)}$ with the cross-sectionally averaged concentration. This further suggests to impose $\langle C^{(n)}\rangle = 0$ for $n>0$. 

\subsubsection{First-order analysis}
At $O(\epsilon)$, the governing equation reads: 
\begin{align}
    \phi\frac{\partial C^{(0)}}{\partial T_1}+\text{Pe}U\frac{\partial C^{(0)}}{\partial Z} = \frac{1}{R}\frac{\partial}{\partial R}\left(R\frac{\partial C^{(1)}}{\partial R }\right)~,\label{eq:C1-dynamics}
\end{align}
together with the boundary condition: 
\begin{align}
    \frac{\partial C^{(1)}}{\partial R}\bigg|_{R=A} = 0~.
\end{align}
Taking the cross-sectional average of the former and using the latter leads to: 
\begin{align}
    \phi\frac{\partial C^{(0)}}{\partial T_1}+\text{Pe}\langle U\rangle\frac{\partial C^{(0)}}{\partial Z} = 0~.\label{eq:C1-averageddynamics}
\end{align}
Combining Eqs.~(\ref{eq:C1-dynamics}) and~(\ref{eq:C1-averageddynamics}), we arrive at the governing equation for $C^{(1)}$, given by:
\begin{align}
\frac{1}{R}\frac{\partial}{\partial R}\left(R\frac{\partial C^{(1)}}{\partial R }\right)= \text{Pe}\left(U-\langle U\rangle\right)\frac{\partial C^{(0)}}{\partial Z} ~.\label{eq:C1-perturbeddynamics}
\end{align}
As such, the axial gradient of $C^{(0)}$ acts as a source term in the linear differential equation governing the radial variation of $C^{(1)}$. Hence, the solution must be in the form:
\begin{align}
    C_{\text{s}}^{(1)} &= \text{Pe}N(R,Z)\frac{\textrm{d} C_{\text{s}}^{(0)}}{\textrm{d} Z}\quad  \text{(steady flow)}\ ,\\
    C_{\text{p}}^{(1)} &= \text{Pe}\mathcal{R}\{M(R,Z,T_1,T_2)\}\frac{\partial C_{\text{p}}^{(0)}}{\partial Z}\quad \text{(oscillatory flow)}\ ,
\end{align}
where the functions $N(R,Z)$ and $M(R,Z,T_1,T_2)$ are to be determined. The perturbative corrections to the cross-sectionally averaged concentration $C^{(0)}$ mentioned above are slaved to its own value.  The strength of the perturbation is controlled by $\textrm{Pe}$ and the functions $N(R,Z)$ and $M(R,Z,T_1,T_2)$ -- these functions quantifying the spatial dependences of the coupling.

For the steady case, substituting the above form into the governing equation for $C^{(1)}$ leads to: 
\begin{align}
    \frac{1}{R}\frac{\partial }{\partial R}\left(R\frac{\partial N}{\partial R}\right) = U_{\text{s}}-\langle U_{\text{s}}\rangle~.
\end{align}
Solving the latter, using Eq.~(\ref{eq:steady-axial-velocity-general}), together with the no-flux boundary condition:
\begin{align}
    \frac{\partial N}{\partial R}\bigg|_{R=A} = 0~,
\end{align}
and the vanishing cross-sectional mean, \textit{i.e.} $\langle N\rangle = 0$, gives:\begin{align}
 N = \langle U_{\text{s}}\rangle\left(\frac{R^2}{4}-\frac{R^4}{8A^2}-\frac{A^2}{12}\right)~.
  \label{nur}
\end{align}

Similarly, for the oscillatory case, one has: 
\begin{align}
       \frac{1}{R}\frac{\partial }{\partial R}\left[R\frac{\partial (\mathcal{R}\{M\})}{\partial R}\right] &= U_\text{p}-\langle U_\text{p}\rangle~.
\end{align}
Solving the latter, using Eq.~(\ref{eq:periodic-axial-velocity-general}), together with the no-flux boundary condition:
\begin{align}
    \frac{\partial M}{\partial R}\bigg|_{R=A} = 0~,
\end{align}
and the vanishing cross-sectional mean, \textit{i.e.} $\langle M\rangle = 0$, gives:
\begin{align}
    M(R,Z,T_1,T_2) = \Gamma \frac{\partial \tilde{P}}{\partial Z}\left[\frac{2}{Aq^3}J_1(qA)-\frac{J_0(qR)}{q^2}+\frac{\beta A^2}{8}-\frac{\beta R^2}{4}\right]~,
\end{align}
where $q = i^{3/2}\text{Wo}$, $\Gamma = \frac{1}{i\text{Wo}^2J_0(qA)}$, and $\beta  = \frac{2J_1(qA)}{qA}$.

\subsubsection{Second-order analysis}
At $O(\epsilon^2)$, the governing equation reads: 
\begin{align}
     \phi\frac{\partial C^{(0)}}{\partial T_2}+\phi\frac{\partial C^{(1)}}{\partial T_1}+\text{Pe}U\frac{\partial C^{(1)}}{\partial Z}+\text{Pe}V\frac{\partial C^{(1)}}{\partial R} = \frac{\partial^2 C^{(0)}}{\partial Z^2}+\frac{1}{R}\frac{\partial}{\partial R}\left(R\frac{\partial C^{(2)}}{\partial R }\right)~,\label{eq:second_order_eqn}
\end{align}
together with the boundary condition: 
\begin{align}
    \frac{\partial C^{(2)}}{\partial R}\bigg|_{R=A}  = \frac{\partial A}{\partial Z}\frac{\partial C^{(0)}}{\partial Z}~.
\end{align}

The cross-sectional average of the second term on the left-hand side of the governing equation is null, since $\langle C^{(1)}\rangle=0$. Besides, cross-sectional averaging the last term on the right-hand side, and invoking the boundary condition, one has: 
\begin{align}
    \left\langle \frac{1}{R}\frac{\partial}{\partial R}\left(R\frac{\partial C^{(2)}}{\partial R }\right)\right\rangle = \frac{2}{A}\frac{\partial A}{\partial Z}\frac{\partial C^{(0)}}{\partial Z}~. \label{eq:C2_avg}
\end{align}
For steady flows, invoking the first-order results, the cross-sectional averaging of the third and fourth terms on the left-hand side of the governing equation leads to:
\begin{align}
    \text{Pe}\left \langle U_\text{s}\frac{\partial C_\text{s}^{(1)}}{\partial Z}\right\rangle&=\text{Pe}^2\left(\langle U_\text{s}N\rangle\frac{\textrm{d}^2 C_\text{s}^{(0)}}{\textrm{d} Z^2}+\left\langle U_\text{s}\frac{\partial N}{\partial Z}\right\rangle\frac{\textrm{d} C_\text{s}^{(0)}}{\textrm{d} Z}\right)~,\\
    \text{Pe}\left \langle V_\text{s}\frac{\partial C_\text{s}^{(1)}}{\partial R}\right\rangle&=\text{Pe}^2\left\langle V_\text{s}\frac{\partial N}{\partial R}\right\rangle\frac{\textrm{d} C_\text{s}^{(0)}}{\textrm{d} Z}~.\label{eq:UC1_VC1_avg}
\end{align}
Similarly, for oscillatory flows, one has:
\begin{align}
    \text{Pe}\left \langle U_\text{p}\frac{\partial C_\text{p}^{(1)}}{\partial Z}\right\rangle &=\text{Pe}^2\left(\langle U_\text{p}\mathcal{R}\{M\}\rangle\frac{\partial^2 C_\text{p}^{(0)}}{\partial Z^2}+\left\langle U_\text{p}\mathcal{R}\left\{\frac{\partial M}{\partial Z}\right\}\right\rangle\frac{\partial C_\text{p}^{(0)}}{\partial Z}\right)~,\\
    \text{Pe}\left \langle V_\text{p}\frac{\partial C_\text{p}^{(1)}}{\partial R}\right\rangle&=\text{Pe}^2\left\langle V_\text{p}\mathcal{R}\left\{\frac{\partial M}{\partial Z}\right\}\right\rangle\frac{\partial C_\text{p}^{(0)}}{\partial Z}~.\label{eq:UpC1_VpC1_avg}
\end{align}
Combining the averages above (Eqs.~(\ref{eq:C2_avg}-\ref{eq:UpC1_VpC1_avg})) with Eq.~(\ref{eq:second_order_eqn}), we obtain the general equation: 
\begin{align}
    \phi\frac{\partial C^{(0)}}{\partial T_2}+(\text{Pe}^2\alpha_\text{s,p}+\alpha_\kappa)\frac{\partial C^{(0)}}{\partial Z} = (1+\text{Pe}^2\gamma_\text{s,p})\frac{\partial^2 C^{(0)}}{\partial Z^2}~,\label{eq:C0-second-order}
\end{align}
where we introduced the auxiliary functions:
\begin{align}
    \alpha_\text{s} &= \left\langle U_\text{s}\frac{\partial N}{\partial Z}\right\rangle+\left\langle V_\text{s}\frac{\partial N}{\partial R}\right\rangle,\\
    \alpha_\text{p} &= \left\langle U_\text{p}\mathcal{R}\left\{\frac{\partial M}{\partial Z}\right\}\right\rangle+\left\langle V_\text{p}\mathcal{R}\left\{\frac{\partial M}{\partial R}\right\}\right\rangle~,\\
    \alpha_\kappa &= -\frac{2}{A}\frac{\partial A}{\partial Z}~,\\
    \gamma_\text{s} &=  -\langle U_\text{s}N\rangle~,\\
    \gamma_\text{p} &= -\langle U_\text{p}\mathcal{R}\{M\}\rangle~.
\end{align}
The auxiliary functions $\alpha_i$ described above quantify the corrections to the advection of the axially-averaged solute concentration. The other auxiliary functions $\gamma_i$ quantify the corresponding change to dispersion. We further note that $\alpha_\text{s}=0$ due to fluid incompressibility, and axial invariance of the cross-sectional fluid flux in the steady case.  

Finally, combining Eqs.~(\ref{t1t2}),~(\ref{eq:C1-averageddynamics}), and~(\ref{eq:C0-second-order})  in a form of heuristic recombination~\cite{roberts1985introduction} leads to the macro-transport equation: 
\begin{align}
    \phi\frac{\partial C^{(0)}}{\partial T}+\left[\text{Pe}\langle U\rangle+\epsilon (\text{Pe}^2\alpha_\text{s,p}+\alpha_{\kappa})\right]\frac{\partial C^{(0)}}{\partial Z} = \epsilon(1+\text{Pe}^2\gamma_\text{s,p})\frac{\partial^2 C^{(0)}}{\partial Z^2}~.
    \label{mactran}
\end{align}
The latter equation is the central result of the present study. It allows to calculate and understand the influence of the flow-induced deformation of the tube walls on Taylor dispersion. Moreover, we note that the corrections to the advection velocity (proportional to the prefactor of $\partial C^{(0)}/\partial Z$) and dispersion coefficient (proportional to the prefactor of $\partial^2 C^{(0)}/\partial Z^2$) are functions of the axial position $Z$ for steady flows, and depend on both $Z$ and $T$ for oscillatory flows.  

\section{Discussion}
\subsection{Steady flow}
We first study the advection velocity and dispersion coefficient in the steady case. Writing the macro-transport equation with dimensional variables, we obtain: 
\begin{align}
    \frac{\partial c_\text{s}}{\partial t}+\left(\langle u_\text{s}\rangle-\frac{2D}{a}\frac{\textrm{d} a}{\textrm{d} z}\right)\frac{\partial c_\text{s}}{\partial z} = D\left(1-\frac{\langle u_\text{s}n\rangle}{D^2}\right)\frac{\partial^2 c_\text{s}}{\partial z^2}~,
\end{align}
where $n=U_0a_0^2N$. The general form of the latter equation has been derived in previous works~\cite{marbach2019active,chang2023taylor} in the context of Taylor dispersion in shaped tubes, using heuristic arguments or invariant-manifold theory. The novelty in our case lies in the coupling to the elastic deformations of the tube walls through the combination of the macro-transport equation with Eqs.~(\ref{wink}),~(\ref{mina}),~(\ref{hydrop}), and~(\ref{nur}). Specifically, here, the advection velocity and dispersion coefficient vary along the axis of the tube because of the elastohydrodynamic modification of the tube radius.

\begin{figure}
     \centering
     \begin{subfigure}[b]{0.3\textwidth}
         \centering
         \includegraphics[width=\textwidth]{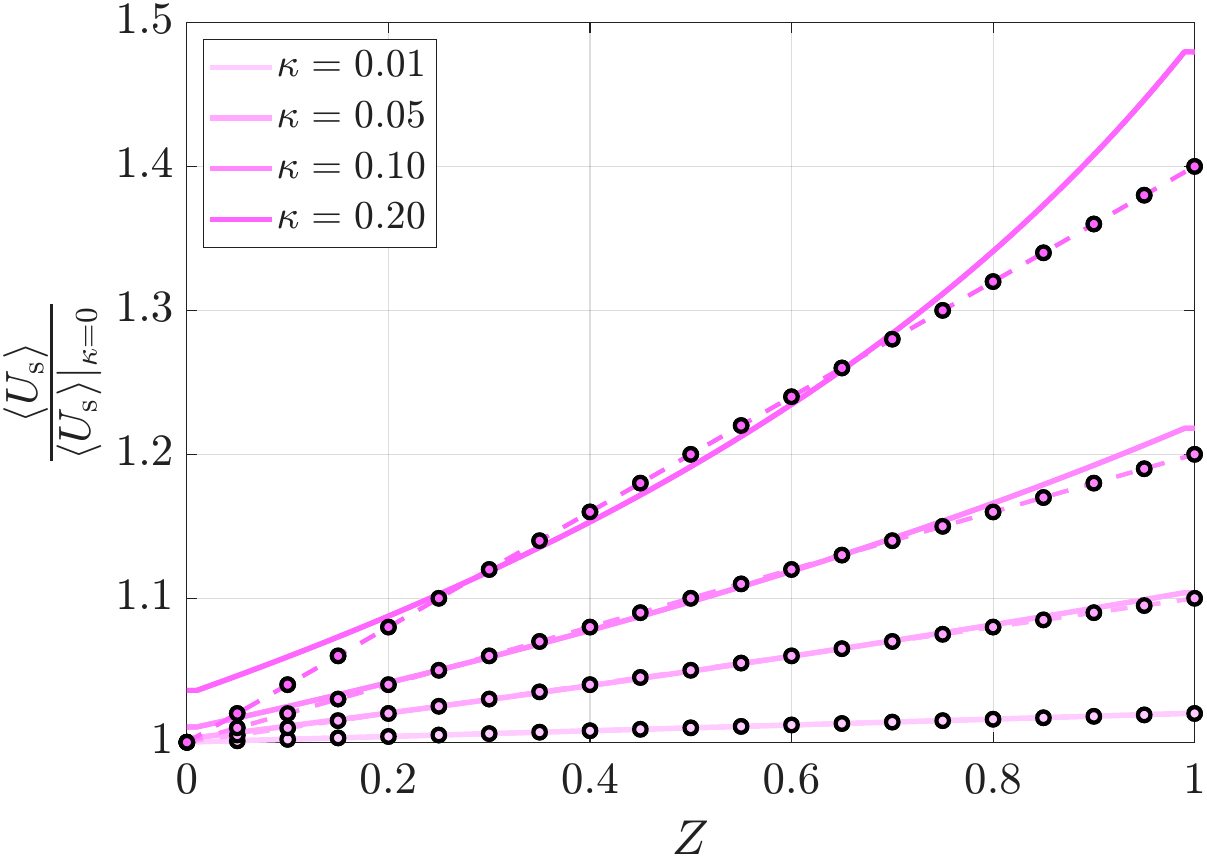}
         \label{fig:2a}
     \end{subfigure}
     \hfill
     \begin{subfigure}[b]{0.28\textwidth}
         \centering
         \includegraphics[width=\textwidth]{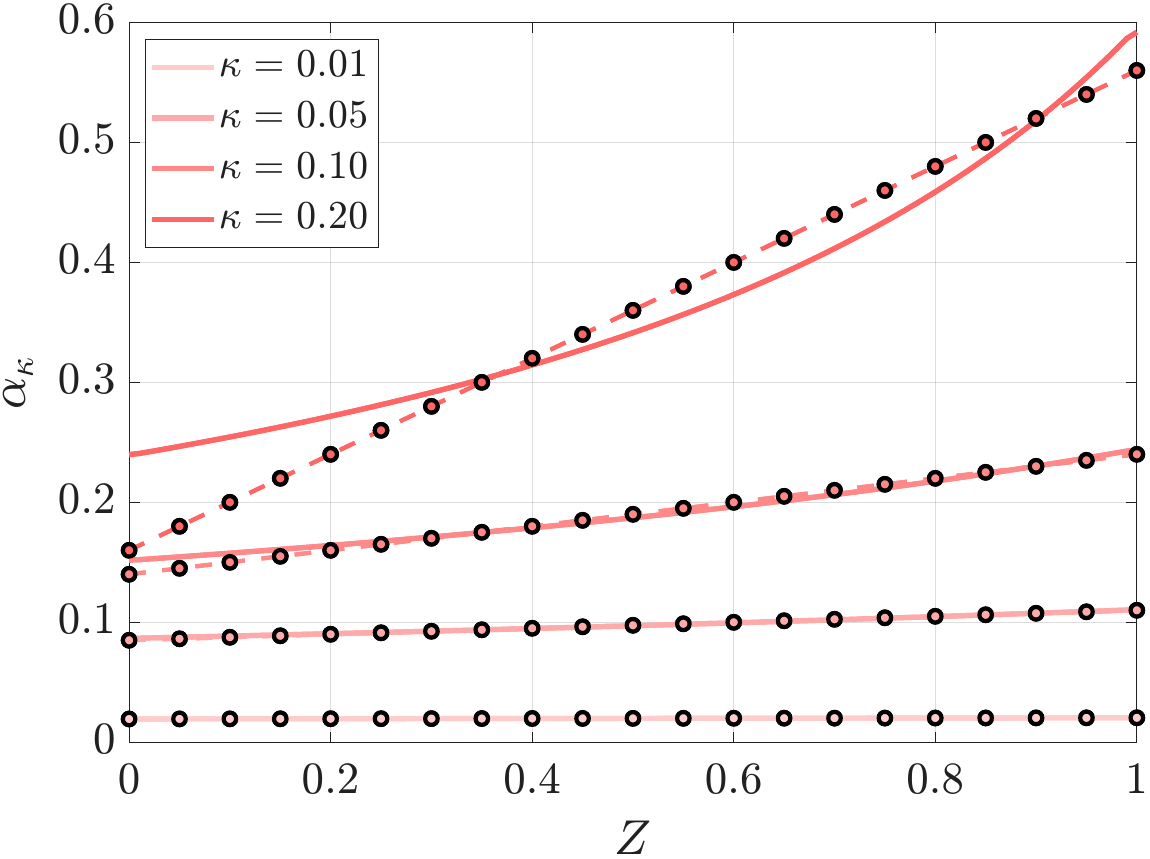}
         \label{fig:2b}
     \end{subfigure}
     \hfill
     \begin{subfigure}[b]{0.3\textwidth}
         \centering
         \includegraphics[width=\textwidth]{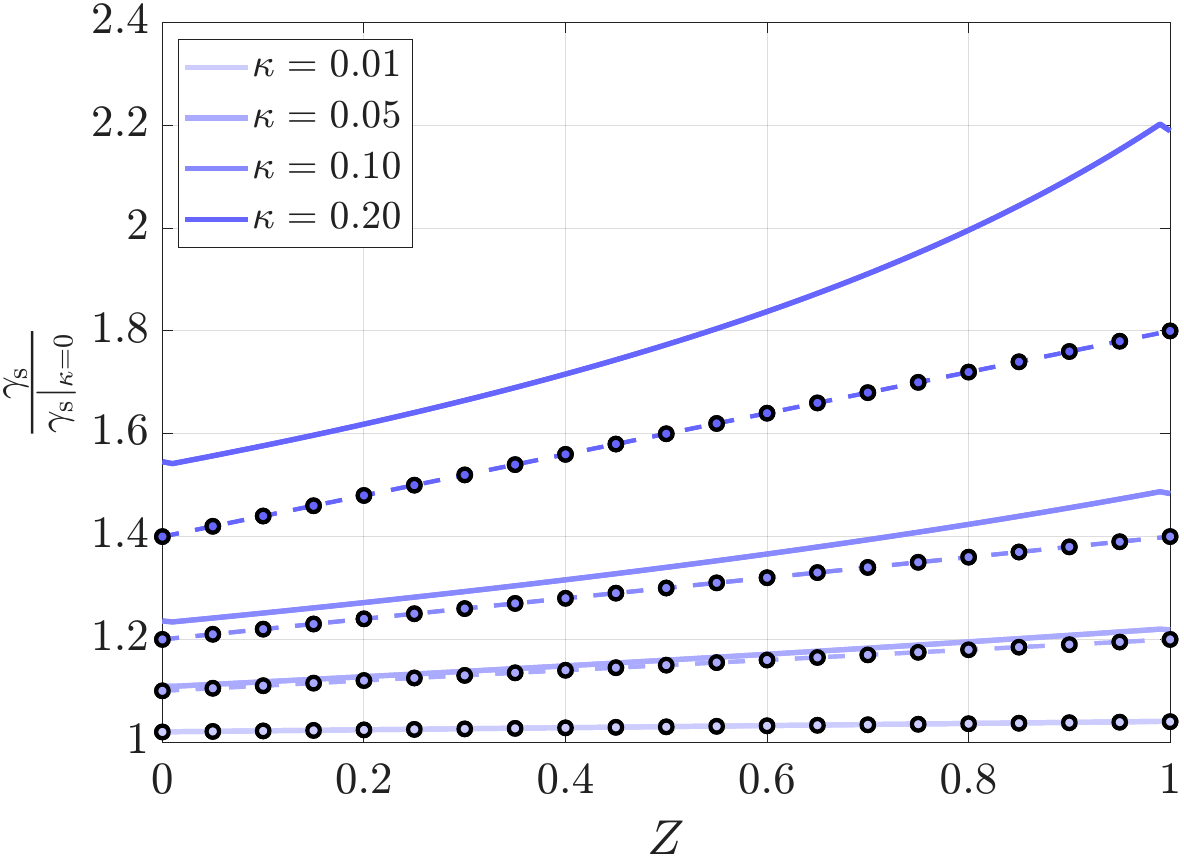}
         \label{fig:2c}
     \end{subfigure}
        \caption{Dimensionless amplification factor $\langle U_\text{s}\rangle/\langle U_\text{s}\rangle|_{\kappa=0}$ of the average flow speed (a), dimensionless control factor $\alpha_\kappa$ of the solutal no-flux contribution to the advection velocity of the solute (b), and solutal-dispersion enhancement factor $\gamma_\text{s}$ (c), as functions of the axial coordinate $Z$ in the tube, and for various values of the dimensionless compliance $\kappa$ of the tube wall, as indicated. The solid lines are obtained from Eqs.~(\ref{e1}),~(\ref{e2}), and~(\ref{e3}), respectively, The dot-dashed lines represent the $O(\kappa^2)$ expansions.}
        \label{fig:steady_flow_correction_factors}
\end{figure}

Let us first study the advection velocity. It is composed of two separate terms that both depend on the elastic compliance of the wall: i) the first term is the classical cross-sectional average of the flow speed, \textit{i.e.} $\langle u_\text{s}\rangle$; ii) the second term, $-(2D/a)(\textrm{d}a/\textrm{d}z)$, is more subtle and results from the solutal no-flux boundary condition at the deformed wall. The dimensionless amplification factor of the average flow speed, with respect to the rigid case ($\kappa\rightarrow0$), reads:
\begin{align}
    \frac{\langle U_\text{s}\rangle}{\langle U_\text{s}\rangle|_{\kappa=0}} = \frac{(\kappa +1)^5-1}{5 \kappa  \left\{(\kappa +1)^5-\left[(\kappa +1)^5-1\right] Z\right\}^{2/5}} \overset{\mathrm{\kappa\rightarrow 0}}{\simeq} 1+2Z\kappa+(1-6Z+7Z^2)\kappa^2+O(\kappa^3)~.
    \label{e1}
\end{align}
Both the full expression and its $O(\kappa^2)$ expansion are plotted in Fig.~\ref{fig:steady_flow_correction_factors}a). We recover the facts that softer walls induce larger average flow speeds and that the effect monotonically increases along the tube axis, due to larger tube obstruction by elastic deformation~\cite{guyard2022elastohydrodynamic}. We now turn to the solutal no-flux contribution to the advection velocity. It is given by the dimensionless expression $\epsilon\alpha_\kappa/\textrm{Pe}$, where:
\begin{align}
    \alpha_\kappa = -\frac{2 \left[1-(\kappa +1)^5\right]}{5 \left\{(\kappa +1)^5-\left[(\kappa +1)^5-1\right] Z\right\}} \overset{\mathrm{\kappa\rightarrow 0}}{\simeq}2 \kappa +2\kappa ^2 (5 Z-3)+O(\kappa^3)~, 
       \label{e2}
\end{align}
which vanishes in the limit of vanishing compliance ($\kappa\rightarrow0$). Both the full expression and its $O(\kappa^2)$ expansion are plotted in Fig.~\ref{fig:steady_flow_correction_factors}b). Interestingly, the solutal no-flux contribution at the deformable wall generates a positive increase in the overall advection velocity. Moreover, the magnitude of the effect grows with the dimensionless compliance $\kappa$. At $O(\kappa)$, the effect is independent of the position $Z$ along the tube axis, and only at $O(\kappa^2)$ does this correction begin to monotonically increase along $Z$. 

Let us now investigate the dispersion part of the problem. The dispersion coefficient takes the Taylor-Aris-like form $D[1+\gamma_\text{s}\text{Pe}^2]$, where $\gamma_\text{s}$ controls the magnitude of the flow-induced amplification of axial diffusion. In particular, when $\kappa=0$, we recover the classical Taylor-Aris result, $\gamma_\text{s}=(1/48)8^{-2}$, for rigid cylindrical tubes~\cite{taylor1953dispersion,aris1956dispersion} (with the additional $8^{-2}$ factor arising out of a different velocity rescaling in our case). For soft tubes, one further has:
\begin{align}
    \frac{\gamma_\text{s}}{\gamma_\text{s}|_{\kappa=0}} = \frac{\left[1-(\kappa +1)^5\right]^2}{25 \kappa ^2 \left\{(\kappa +1)^5-\left[(\kappa +1)^5-1\right] Z\right\}^{2/5}}\overset{\mathrm{\kappa\rightarrow 0}}{\simeq}1+2(1+Z) \kappa  +\kappa ^2 \left(7Z^2-2Z+3\right)+O(\kappa^3)~,
       \label{e3}
\end{align}
with a position-dependent linear correction to the dispersion. Both the full expression and its $O(\kappa^2)$ expansion are plotted in Fig.~\ref{fig:steady_flow_correction_factors}c). Interestingly, softness generates a positive increase in the dispersion coefficient. Moreover, the magnitude of the effect grows with the dimensionless compliance $\kappa$ and the axial coordinate $Z$. This central result may seem contradictory with the classical Taylor-Aris picture at first sight, since the effective tube radius reduces due to the increasing elastic deformation~\cite{guyard2022elastohydrodynamic} (and thus reduces the local $\textrm{Pe}$), but is in fact fully consistent with the imposed-pressure configuration and the steady fluid-flux conservation along the tube axis.

\subsection{Oscillatory flow}
Here, we study the advection velocity and dispersion coefficient in the oscillatory case. As in the steady case, they vary along the axis of the tube, but they also depend on time now. 

\begin{figure}
     \centering
     \begin{subfigure}[b]{0.45\textwidth}
         \centering
         \includegraphics[width=\textwidth]{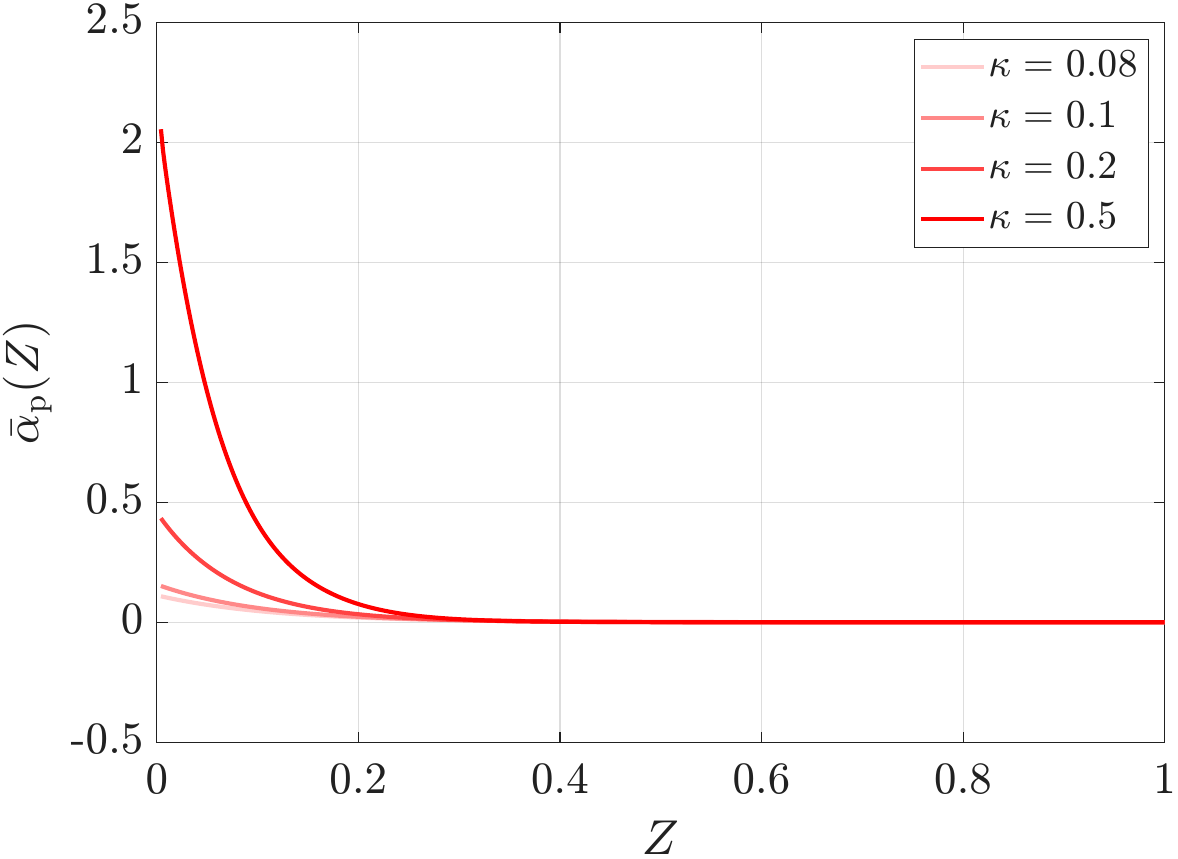}
         \label{fig:5a}
     \end{subfigure}
     \hfill
     \begin{subfigure}[b]{0.45\textwidth}
         \centering
         \includegraphics[width=\textwidth]{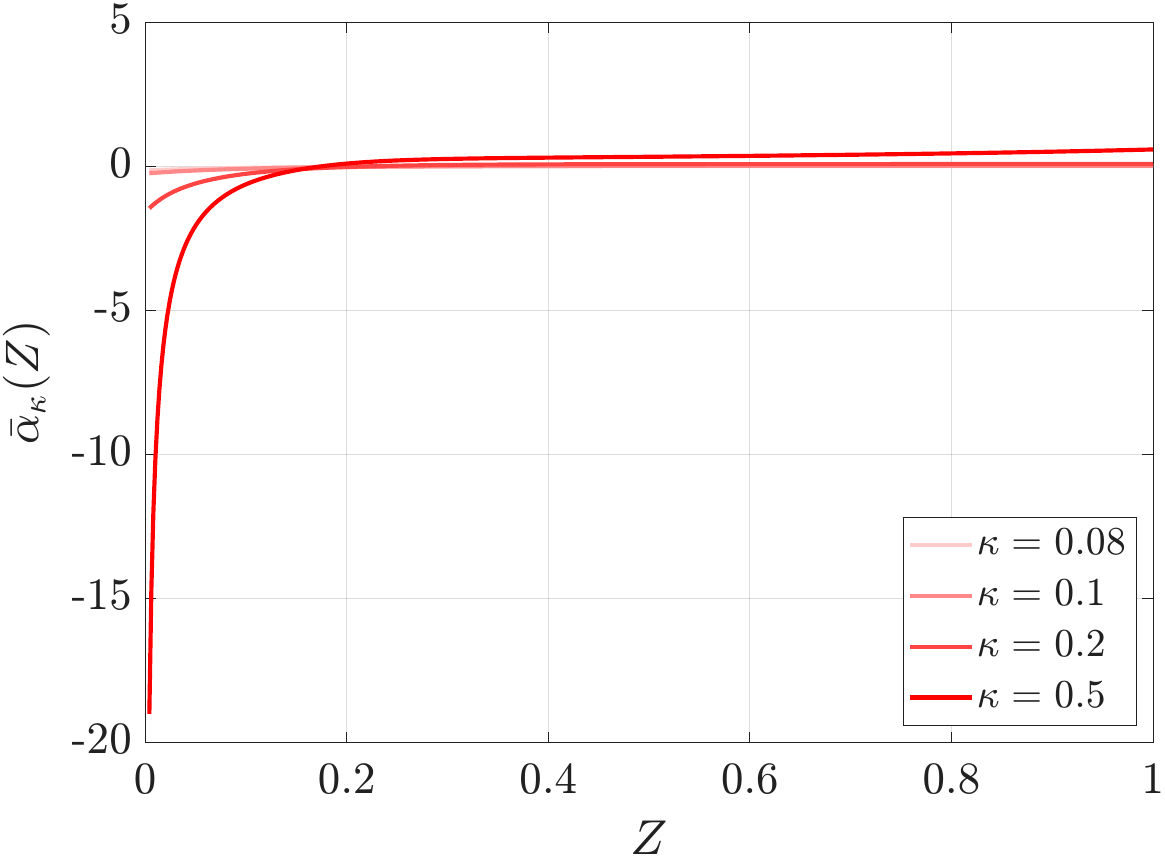}
         \label{fig:5b}
     \end{subfigure}
        \caption{Numerically-computed values of the two dimensionless factors $\bar{\alpha}_\text{p}$ (a) and $\bar{\alpha}_\kappa$ (b) controlling the period-averaged advection velocity of the solute, as functions of the axial coordinate $Z$ in the tube, for $\text{Wo} = 0.7$ and $\text{St} = 5$, and for various values of the dimensionless compliance $\kappa$ of the tube wall, as indicated.}
        \label{fig:advection_kappa_coefficient_oscillatory}
\end{figure}

Let us first study the advection velocity. In the oscillatory case, one has $\bar{U}_\text{p}=0$, where $\bar{\cdot}$ denotes the average over one oscillation period. The average advection velocity is then composed of two separate terms that both depend on the elastic compliance of the wall, and are controlled by the two dimensionless factors $\bar{\alpha}_\text{p}$ and $\bar{\alpha}_\kappa$. Their numerical evaluations are shown in Fig.~\ref{fig:advection_kappa_coefficient_oscillatory}. We observe that $\bar{\alpha}_\text{p}$ is vanishingly small at vanishing compliance. As the compliance increases, the overall magnitude of both factors increases, $\bar{\alpha}_\text{p}$ being positive and monotonically decaying with $Z$, and $\bar{\alpha}_\kappa$ changing sign, from negative to positive, as $Z$ increases. 

\begin{figure}
     \centering
     \begin{subfigure}[b]{0.45\textwidth}
         \centering
         \includegraphics[width=\textwidth]{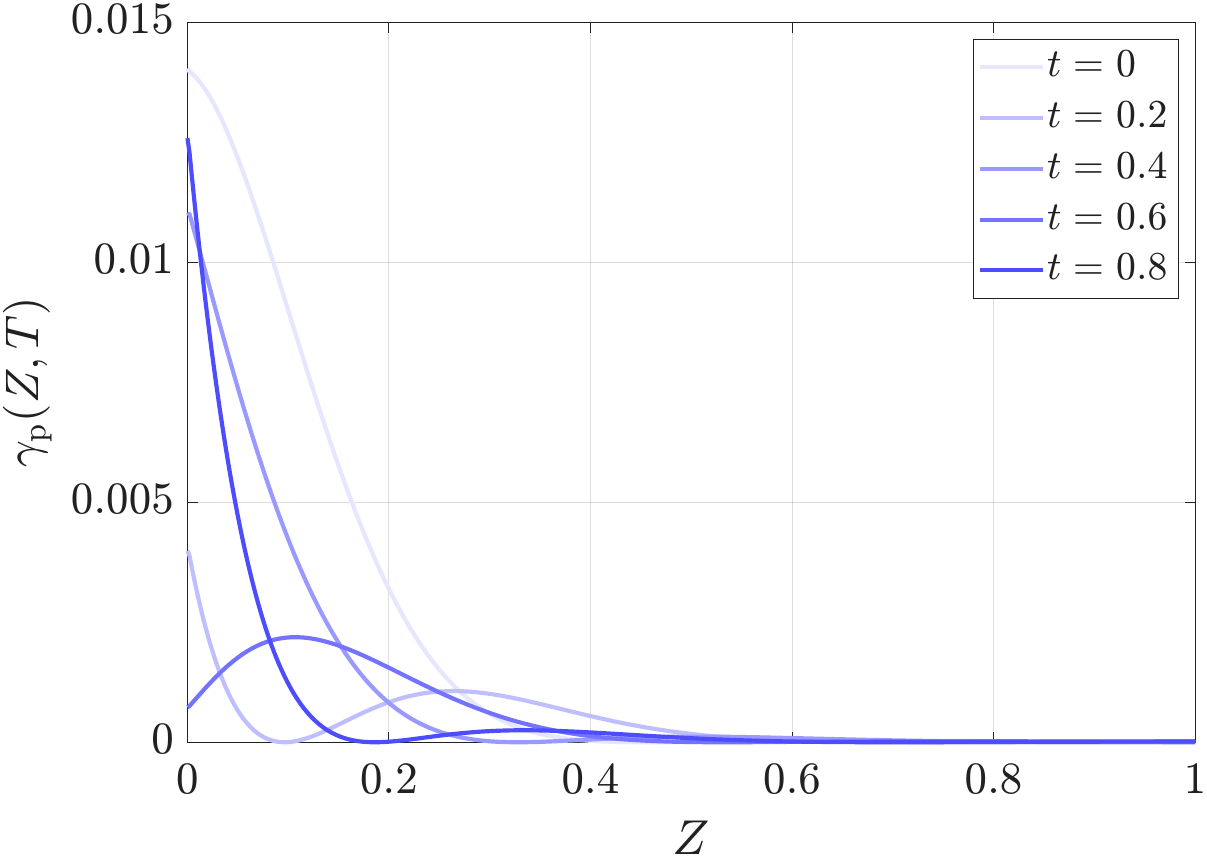}
         \label{fig:4a}
     \end{subfigure}
     \hfill
     \begin{subfigure}[b]{0.45\textwidth}
         \centering
         \includegraphics[width=\textwidth]{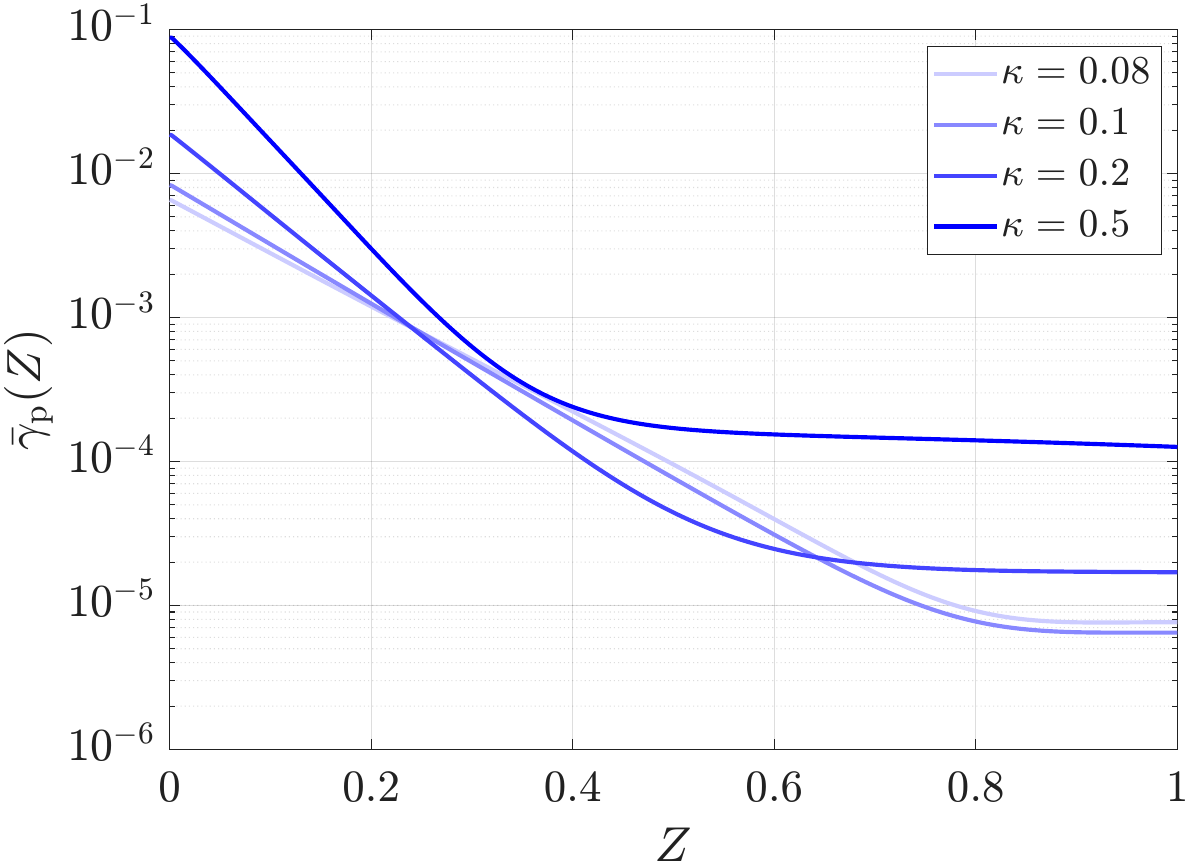}
         \label{fig:4b}
     \end{subfigure}
        \caption{a) Numerically-computed value of the solutal-dispersion enhancement factor $\gamma_\text{p}$, as a function of the axial coordinate $Z$ in the tube, for $\text{Wo} = 0.7$, $\text{St} = 5$ and $\kappa = 0.08$, and for various dimensionless times $T$ within one oscillation period, as indicated. b) Corresponding period-averaged solutal-dispersion enhancement factor $\bar{\gamma}_\text{p}$ as a function of the axial coordinate $Z$ in the tube, for $\text{Wo} = 0.7$ and $\text{St} = 5$, and for various values of the dimensionless compliance $\kappa$ of the tube wall, as indicated.}
        \label{fig:dispersion_coefficient_oscillatory}
\end{figure}

Let us now investigate the dispersion part of the problem. The dispersion coefficient 
is controlled by $\gamma_\text{p}$. Its numerical evaluation is displayed in Fig.~\ref{fig:dispersion_coefficient_oscillatory}, for both $\gamma_\text{p}$ and its period-averaged value, and only shows positive values, which indicates a systematic enhancement of solutal dispersion by the flow, as in classical Taylor-Aris dispersion for a rigid cylindrical tube~\cite{taylor1953dispersion,aris1956dispersion}. In contrast to the latter, dispersion becomes inhomogeneous and complex in the soft case. In particular, most of the effect seems to occur near the tube inlet ($Z=0$). It is important to recall here the elastohydrodynamic mode-coupling effect mentioned at the end of Sec.~\ref{floprof}. This feature and its consequences can be further understood by performing a perturbation expansion at small compliance, as provided in the Appendix, where the $O(\kappa)$ pressure field is shown to oscillate with two different frequencies. Since one has $\gamma_\text{p}\sim (\partial P/\partial Z)^2$, these modes further interact together and generate more harmonics, hence producing a complex overall temporal behaviour. We also highlight here that this entire discussion ignores the effects of convective inertia, which is only true for $\epsilon \textrm{Re} \ll \kappa$. The latter constraint limits the minimum value of $\kappa$ chosen in the figures since $\epsilon \textrm{Re} = \textrm{Wo}^2/(2\pi\textrm{St})$.

\subsection{Implications for solute dispersion}
\begin{figure}[h]    
        \includegraphics[angle = 0,width=\textwidth]{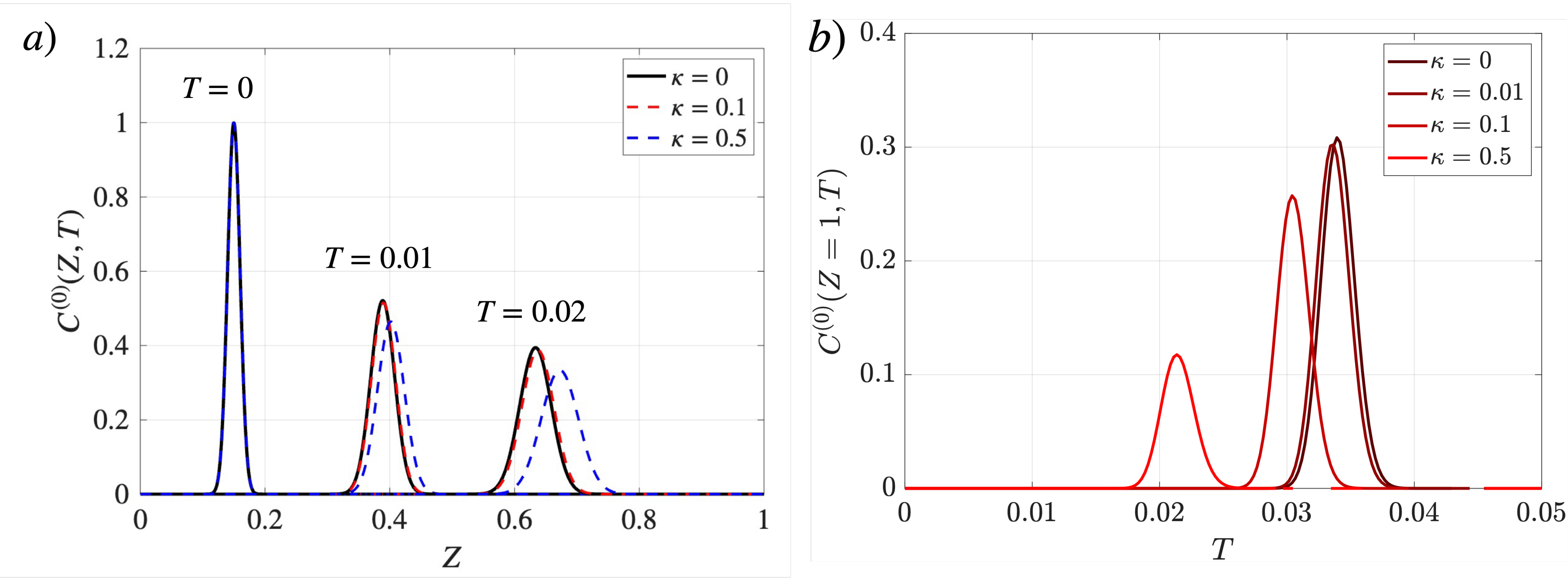}
        \caption{Dimensionless solute concentration profile $C^{(0)}$ as a function of axial coordinate $Z$, for three different dimensionless times $T$ and two values of the dimensionless compliance $\kappa$, as indicated, as obtained from numerically solving Eq.~(\ref{mactran}) with $\phi=1$, $\text{Pe} = 200$, $\epsilon = 0.001$, and using the steady-flow parameters. 
        The initial condition is a Gaussian distribution localised at $Z = 0.15$ with a standard deviation $\sigma = 1/(10\sqrt{50})$. At the tube inlet we apply a full no-flux condition ($UC^{(0)}-\partial C^{(0)}/\partial Z = 0$), including both advection and diffusive fluxes, and at the tube outlet we impose $\partial C^{(0)}/\partial Z=0$ to allow for the effective solute removal by the flow. b) Dimensionless solute concentration profile $C^{(0)}$ at the tube outlet ($Z=1$) as a function of the dimensionless time $T$, for various values of the dimensionless compliances $\kappa$ as indicated.}
        \label{fig:solute_dispersion_tube}
\end{figure}

Lastly, we wish to illustrate some practical implications of tube elasticity on solutal dispersion. To do so, we focus on the steady-flow case and compute the solute concentration profile $C^{(0)}(Z,T)$ over space and time by solving Eq.~(\ref{mactran}). Note that we are allowed to set $\phi=1$ here, despite the fact that $\omega=0$ in the steady-flow case. Indeed, a change of time scale in the non-dimensionalization procedure has no influence on the steady flow part of the problem (since $\phi \frac{\partial}{\partial T} = \frac{la_0}{D}\frac{\partial}{\partial t}$, which vanishes under steady flow). The results are shown in Fig.~\ref{fig:solute_dispersion_tube} and exhibit two main features, in agreement with the above findings. First, the softer the tube, the larger the advection velocity, and thus the earlier the outlet is reached by the concentration peak. Second, the solute disperses more with increasing compliance. Besides the two main points above, an additional feature is that the fore-aft asymmetry in the temporal domain is enhanced at larger compliance.
\section{Conclusion}
We investigated theoretically and numerically the Taylor-Aris dispersion in a soft tube, the latter being described by a Winkler elastic model. In particular, using multiple-time-scale analysis, we derived the characteristic macro-transport equation for cross-sectionally-averaged solutal transport, in both steady and oscillatory flows. Our work exhibits the modifications induced by wall elasticity to effective advection velocity and dispersion coefficient. Mainly, softness enhances both quantities. This could have direct implications in biology and microfluidic technologies. Beyond these findings, the simulated solutal dynamics in a soft tube offers the possibility to infer the compliance of the tube itself in practice by measuring the distribution of the solute at a given point. This suggests a new non-invasive methodology for the inference of elastic properties of the tube based on standard flow analysis. Specifically, for biomedical applications, this could lead to the development of detection tools for characterising sudden weakening in blood vessels~\cite{han2012twisted}, that is one of the major causes of heart diseases. Finally, our work points to several natural extensions. An important next step is to look at diffusio-osmotic and diffusio-phoretic flows in soft tubes, where the solute dynamics drives the flow leading to a non-trivial two-way coupling of the elastohydrodynamics and the solute transport. Finally, recent studies on dispersion of active particles have stressed the role of confinement and the current work highlights how wall elasticity could provide a control parameter for the active Taylor-Aris dispersion. 

\section{Conflicts of interest}
There are no conflicts of interest to declare.

\section{acknowledgments}
The authors thank Alexandre Vilquin and Anthony Robert for interesting discussions. A.J. thanks the Herchel Smith Fund for a Postdoctoral Fellowship. The authors acknowledge financial support from the Agence Nationale de la Recherche under Softer (ANR21-CE06-0029) and Fricolas (ANR-21-CE06-0039) grants, as well as from the Interdisciplinary and Exploratory Research Program under MISTIC grant at the University of Bordeaux, France. The authors also acknowledge financial support from the European Union through the European Research Council under EMetBrown (ERC-CoG-101039103) grant and NoDiCE (ERC-2024-COG-101170653). Views and opinions expressed are however those of the authors only and do not necessarily reflect those of the European Union or the European Research Council. Neither the European Union nor the granting authority can be held responsible for them. Finally, they thank the RRI Frontiers of Life, which received financial support from the French government in the framework of the University of Bordeaux's France 2030 program, as well as the Soft Matter Collaborative Research Unit, Frontier Research Center for Advanced Material and Life Science, Faculty of Advanced Life Science, Hokkaido University, Sapporo, Japan, and the CNRS International Research Network between France and India on ``Hydrodynamics at small scales: from soft matter to bioengineering".


\newpage
\section*{APPENDIX A: Oscillatory flow in a weakly deformable tube}
For a weakly-deformable tube, we approach the problem with a perturbation expansion in $\kappa$, through:
\begin{align}
    \tilde{P} = \tilde{P}_0+\kappa \tilde{P}_1+O(\kappa^2),
\end{align}
as has been done in previous works~\cite{pande2023oscillatory,anand2020transient,rade2025theory,zhang2024elasto}. At $O(1)$, one has:
\begin{align}
    \frac{\textrm{d}^2\tilde{P}_0}{\textrm{d}Z^2}=0~,
\end{align}
together with the boundary conditions:
\begin{align}
    \tilde{P}_0(Z = 0,T) &= \textrm{e}^{2\pi iT},\\
    \tilde{P}_0(Z =1,T) &= 0,
\end{align}
which lead to: 
\begin{align}
    \tilde{P}_0 = (1-Z)\textrm{e}^{2\pi iT}~.
\end{align}
At $O(\kappa)$, one has: 
\begin{align}
    G_0(\text{Wo})\frac{\partial^2\tilde{P}_1}{\partial Z^2} = -\frac{\partial}{\partial Z}\left[G_1(\text{Wo})\mathcal{R}\{\tilde{P}_0\}\frac{\partial \tilde{P}_0}{\partial Z}\right]+2\,\text{St}\frac{\partial \tilde{P}_0}{\partial T}~,
\end{align}
where:
\begin{align}
    G_0(\text{Wo}) & = \frac{1}{2i\text{Wo}^2}\frac{J_2(i^{3/2}\text{Wo})}{J_0(i^{3/2}\text{Wo})}~,\\
    G_1(\text{Wo}) &= \frac{i}{\text{Wo}^2}\frac{J_1(i^{3/2}\text{Wo})^2}{J_0(i^{3/2}\text{Wo})^2}~. 
\end{align}
All together, these lead to:
\begin{align}
    \tilde{P}_1(Z,T) = \frac{1}{6}Z(1-Z)\left[3\frac{G_0(\text{Wo})}{G_1(\text{Wo})}\mathcal{R}\{\textrm{e}^{2\pi i T}\}\textrm{e}^{2\pi i T}+(Z-2)\frac{\text{St}}{G_0(\text{Wo})}2\pi i \textrm{e}^{2\pi i T}\right]~.
\end{align}
The real, total pressure field, up to $O(\kappa)$, is then given by $\mathcal{R}\{   \tilde{P}_0(Z,T)\}+\kappa \mathcal{R}\{   \tilde{P}_1(Z,T)\}$, where harmonic generation is apparent. In addition, the average of the pressure field over one oscillation period does not vanish at $O(\kappa)$~\cite{zhang2024elasto,pande2023oscillatory,rade2025theory}, and thus generates a streaming flow in the tube. It is important to highlight here that this streaming flow ignores convective inertial effects. Previous work~\cite{zhang2024elasto} has shown that this leads to a deviation in the net streaming flow, unless $\epsilon \textrm{Re}\ll \kappa$.  In our study, we stick to $\epsilon \textrm{Re}\ll \kappa$. However, further work needs to characterise the inclusion of convective inertial effects.

\bibliography{Jha2026}
\bibliographystyle{vancouver}
\newpage

\end{document}